\documentclass[twocolumn,times,final,authoryear]{elsarticle}

\usepackage{prletters-preprint}
 
\usepackage{framed,multirow}

\usepackage{amssymb}
\usepackage{latexsym}
\usepackage{amsmath}
\usepackage{url}
\usepackage[dvipsnames]{xcolor}
\definecolor{newcolor}{rgb}{.8,.349,.1}

\usepackage[framemethod=TikZ]{mdframed}
\definecolor{framecolor}{RGB}{0,0.541,0.855}

\usepackage[breaklinks,colorlinks]{hyperref}
\usepackage{booktabs}
\usepackage{subcaption}
\usepackage{multirow}
\usepackage{pgfplots}
\usepackage{amsmath}

\newcommand{\argmin}{\mathop{\mathrm{arg\,min}}}

\newcommand{\std}[1]{\scriptsize{$\pm$#1}}

\newcommand{\rebuttal}[1]{#1}
\newcommand{\rebuttalrev}[1]{#1}

\journal{Pattern Recognition Letters}

\begin{document}

\setcounter{page}{1}

\begin{frontmatter}

\title{Robust brain age estimation from structural MRI with contrastive learning}

\author[unito]{Carlo Alberto \surname{Barbano}\corref{cor1}} 
\cortext[cor1]{Corresponding author.}
\ead{carlo.barbano@unito.it}

\author[neurospin]{Benoit \surname{Dufumier}}
\ead{benoit.dufumier@cea.fr}

\author[neurospin]{Edouard \surname{Duchesnay}}
\ead{edouard.duchesnay@cea.fr}

\author[unito]{Marco \surname{Grangetto}}
\ead{marco.grangetto@unito.it}

\author[telecom]{Pietro \surname{Gori}}
\ead{pietro.gori@telecom-paris.fr}

\author{for the Alzheimer's Disease Neuroimaging Initiative}\fnref{cor2}
\fntext[cor2]{Data used in preparation of this article were obtained from the Alzheimer's Disease Neuroimaging Initiative (ADNI) database (adni.loni.usc.edu). As such, the investigators within the ADNI contributed to the design and implementation of ADNI and/or provided data but did not participate in the analysis or writing of this report. A complete listing of ADNI investigators can be found at: \url{http://adni.loni.usc.edu/wp-content/uploads/how_to_apply/ADNI_Acknowledgement_List.pdf}}

\affiliation[unito]{organization={University of Turin},
                city={Turin}, 
                postcode={10149}, 
                country={Italy}}

\affiliation[neurospin]{organization={BAOBAB, NeuroSpin, CEA},
                city={Saclay}, 
                citysep={},
                postcode={91191}, 
                country={France}}

\affiliation[telecom]{
    organization={LTCI, Télécom Paris, IP Paris},
    city={Palaiseau},
    postcode={91120},
    country={France}
}

\begin{abstract}
Estimating brain age from structural MRI has emerged as a powerful tool for characterizing normative and pathological aging. In this work, we explore contrastive learning as a scalable and robust alternative to \rebuttalrev{L1-}supervised approaches for brain age estimation. We introduce a novel contrastive loss function, $\mathcal{L}^{exp}$, and evaluate it across multiple public neuroimaging datasets comprising over 20,000 scans. Our experiments reveal four key findings. First, scaling pre-training on diverse, multi-site data consistently improves generalization performance, cutting external mean absolute error (MAE) nearly in half. Second, $\mathcal{L}^{exp}$ is robust to site-related confounds, maintaining low scanner-predictability as training size increases. Third, contrastive models reliably capture accelerated aging in patients with cognitive impairment and Alzheimer's disease, as shown through brain age gap analysis, ROC curves, and longitudinal trends. Lastly, unlike \rebuttalrev{L1-}supervised baselines, $\mathcal{L}^{exp}$ maintains a strong correlation between brain age accuracy and downstream diagnostic performance, supporting its potential as a foundation model for neuroimaging. These results position contrastive learning as a promising direction for building generalizable and clinically meaningful brain representations.
\end{abstract}

\end{frontmatter}

\section{Introduction}

Brain aging is a complex and individualized process, involving structural changes such as cortical thinning and white matter loss, which vary widely across the population. Accurately modeling these subject-specific trajectories is a long-standing challenge in neuroscience, with high clinical relevance for detecting and tracking age-related neurodegenerative diseases. Recent Deep Learning (DL) approaches have shown promise in building brain-age predictors that learn mappings from anatomical MRI to chronological age~\citep{peng2021accurate}. However, these models typically require large-scale training data, which necessitates the aggregation of scans from multiple imaging sites and studies.
A major challenge with multi-site data is that DL models, especially deep neural networks, tend to overfit to site-specific artifacts, such as differences in scanner hardware, acquisition protocols, or magnetic field strength. This phenomenon leads to degraded performance on unseen sites and severely limits real-world applicability~\citep{glocker2019machine,wachinger2021detect}. As a result, the development of models that generalize well across heterogeneous sources remains an open and pressing problem.
The OpenBHB challenge~\citep{dufumier2022openbhb} was introduced as a benchmark to address this issue, offering a large-scale, multi-site dataset designed to evaluate brain-age prediction under real-world generalization constraints. Traditional models participating in this benchmark have largely relied on \rebuttalrev{standard} supervised regression objectives such as L1~\citep{cole2017predicting, jonsson2019brain,peng2021accurate}. However, these frameworks typically ignore the presence of site-specific confounds during training, which may affect the learned representations.
To address this, we previously introduced a Contrastive Learning (CL) framework for brain-age regression~\citep{barbano2023contrastive}, based on a novel similarity-driven loss function, $\mathcal{L}^{exp}$, which encourages biologically meaningful and site-invariant embeddings. In this extended work, we scale up our previous approach both in data and scope: we employ over 20,000 structural T1 scans from four large public datasets (OpenBHB, ADNI, ICBM, and OASIS-3), and conduct a broad empirical study to analyze generalization, site robustness, clinical validity, and downstream transferability.
Our contributions are as follows:

\begin{itemize}
    \item We significantly improve the state-of-the-art performance in brain age prediction on the OpenBHB challenge;
    \item We show that our proposed formulation $\mathcal{L}^{exp}$ achieves consistent invariance towards site bias;
    \item We show that CL is a reliable estimator of accelerated brain aging in patients with cognitive impairment (MCI) and Alzheimer’s disease (AD);
    \item We show that better brain age estimators based on CL can result in better diagnostic models.
\end{itemize}

\section{Related Works}

Our contribution is based on the related literature in CL, metric learning, fairness, and debiasing. 

\paragraph{Contrastive learning} %
When label information is not available, instance-based discrimination approaches~\citep{chen2020simple} have gained considerable success, almost closing the gap with its supervised counterpart. Most recent research in CL has focused on designing the right data augmentations~\citep{chen2020simple,zhang2022rethinking}, contrastive losses~\citep{chen2020simple, caron2021emerging,zbontar2021barlow,grill2020bootstrap}, memory banks~\citep{he2020momentum,dwibedi2021little}, prototypes (or clusters)~\citep{caron2020unsupervised,li2020prototypical} typically for downstream classification, detection or segmentation tasks, leaving open the question for regression tasks. When weak labels are available (either discrete or continuous), kernel-based approaches have been explored for CL~\citep{dufumier2021contrastive, dufumier2023integrating, sarfati_learning_2023, dufumier_conditional_2021}, using a re-weighting scheme to adapt the alignment term between samples with close weak labels. When continuous labels are available, ranking-based approaches~\citep{zha2023rank} have recently gained traction by showing strong generalization results on regression tasks. \rebuttal{Also recently, adaptive neighborhood selection in contrastive learning has been proposed~\citep{trauble2024contrastive,traeble2024adaptive} in the context of brain age regression using Magnetic Resonance Elastography (MRE) and demonstrated good generalization results for non-uniformly age distributed datasets.}

\paragraph{Brain age estimation} Estimating the chronological age from brain imaging in healthy individuals has been a challenge for machine learning models for the past 10 years. The first traditional models (linear, SVM) were trained on handcrafted features extracted from structural MRI (cortical thickness, gray matter volume, surface area)~\citep{cole2015prediction,franke2012longitudinal}. Then, more complex deep learning models, such as CNNs~\citep{cole2017predicting} or Transformers~\citep{he2021global}, have been deployed to automatically extract brain features from raw imaging. These brain age models were used to derive a brain age gap correlated to common brain disorders such as schizophrenia, bipolar disorder, ASD, and ADHD~\citep{kaufmann2019common} and heritable~\citep{cole2017predicting,jonsson2019brain}. Nevertheless, recent studies~\citep{dufumier2024exploring} showed that DL models generalize poorly to new MRI sequences/acquisitions, and they still underperform on populations with different demographic characteristics. \rebuttal{Additionally, it was shown that more recent Transformer-based architectures do not improve brain age models compared to classical CNNs, and they seem to extract similar concepts in their representations~\citep{siegel2025transformers}. Finally, regarding the clinical utility of brain age models, a recent study~\citep{tan2025mind} suggests that directly using the brain age gap as a biomarker for Alzheimer's disease prediction might be suboptimal compared to models directly fitted for this task, in line with other studies~\citep{schulz2024beyond,bashyam2020mri} suggesting that tightly-fitted brain age models do not necessarily transfer well on clinical tasks.}

\begin{table}
    \centering
    \caption{Study cohorts composition.}
    \resizebox{\linewidth}{!}{%
    \begin{tabular}{c c c c c}
        \toprule
        & N. Subjects & N. Images & Age, years & Age range, years \\
        \midrule
        OpenBHB &  3984 & 3984 & 24.92 (14.29) & [5.9, 88] \\
        ADNI & 1754 & 14332 & 76.22 (7.17) & [50.5, 97.4] \\
        ICBM & 622 & 977 & 33.71 (13.94) & [18, 80] \\
        OASIS3 & 685 & 1354 & 67.77 (9.16) & [42, 95] \\
        \bottomrule
    \end{tabular}
    }
    \label{tab:datasets}
\end{table}

\begin{table}
    \centering
    \caption{Pretrain size (healthy subjects)}
    \label{tab:pretrain-healthy-size}
    \resizebox{\linewidth}{!}{%
    \begin{tabular}{c c c c c c}
    \toprule
            & OpenBHB & ADNI & ICBM & OASIS-3 & Total\\
    \midrule
    Number of images &  3984 & 4367 & 977 & 1262 & 10590\\
    Number of subjects & 3984 & 633 & 622 & 597 & 5836 \\
    \bottomrule
    \end{tabular}
    }
\end{table}

\section{Materials and Method}

\subsection{Experimental data}

We use a collection of T1-weighted MRI scans comprising 7,045 different individuals, for a total of 20,647 images. The experimental data is gathered from different publicly available datasets: OpenBHB, ADNI, ICBM, and OASIS. The overall composition of the cohorts used in this study is presented in Tab.~\ref{tab:datasets}. %
OpenBHB~\citep{dufumier2022openbhb} is a recently released dataset %
that aggregates healthy control (HC) samples from many public cohorts (ABIDE 1, ABIDE 2, CoRR, GSP, IXI, Localizer, MPI-Leipzig, NAR, NPC, RBP). Every scan comes from a different subject. %
We employ all the publicly available training data (N=3984). For testing, OpenBHB relies on two private test sets, one internal (with the same acquisition sites as the public training set), and one external (with different ones). \rebuttal{It is worth noting that, thanks to the number of datasets gathered, OpenBHB provides one of the widest diversity in terms of population in neuroimaging data, including American, European, and Asian subjects.}
For ADNI~\citep{petersen2010alzheimer}\footnote{Data used in the preparation of this article were obtained from the Alzheimer's Disease Neuroimaging Initiative (ADNI) database (\url{adni.loni.usc.edu}).}
, we include all phases ADNI-1, ADNI-2, ADNI-GO, and ADNI-3 in our study, amounting to 633 HC, 712 MCI (483 sMCI, 229 pMCI), and 409 AD patients. For ICBM~\citep{mazziotta2001probabilistic}\footnote{Data used in the preparation of this work were obtained from the International Consortium for Brain Mapping (ICBM) database (\url{www.loni.usc.edu/ICBM}).} 
and OASIS3~\citep{lamontagne2019oasis}, we include 622 and 685 subjects respectively, with the latter containing 88 AD %
cases. %

\paragraph{Image acquisition and preprocessing}
All images underwent the same standard VBM preprocessing, using CAT12, which includes non-linear registration to the MNI template and gray matter (GM) extraction. The final spatial resolution is 1.5mm isotropic and the images are of size 121 x 145 x 121.
OpenBHB provides T1-weighted MRI scans acquired with 1.5T and 3T scanner. The data is already preprocessed with CAT12 and gray-matter volume is provided for download. 
For ADNI, we downloaded original T1-weighted images acquired with 1.5T and 3T scanners in the DICOM format, and we preprocessed them with the same pipeline of OpenBHB, using the brainprep package\footnote{\url{https://github.com/neurospin-deepinsight/brainprep}}, which is based on CAT12, in order to extract GM volumes. We included MRI scans acquired with the MP-RAGE protocol. For further details about the acquisition, we refer to the official documentation.\footnote{\url{https://adni.loni.usc.edu/methods/mri-tool/mri-analysis/}}.
ICBM and OASIS3 provide T1-weighted MRI scans acquired with 1.5T and 3T scanners, and we preprocessed them with the same pipeline.

\subsection{Experimental setup}

\paragraph{Experimental protocol} 
Across all of the experiments we employ a 5-fold CV scheme, at the subject level, where the splits are stratified by age w.r.t to the original distribution. %
In experiments targeting only brain age prediction, we consider only HC patients (the distribution is presented in Tab.~\ref{tab:pretrain-healthy-size}), whereas in experiments targeting cognitive decline we consider all patients (HC, AD, and MCI). %
As network architecture, we employ the 3D implementations of ResNet-18 with 33.2M parameters, as previous works~\rebuttalrev{\citep{dufumier2021benchmarking, dufumier2022openbhb,dufumier2024exploring,barbano2023contrastive,barbano2025anatcl,siegel2025transformers,yang2025ensemble}} show that \rebuttalrev{different model sizes and architectures} do not provide better results. %
\rebuttal{In addition to our proposed losses, we also perform experiments with RNC~\citep{zha2023rank} and AdapN~\citep{trauble2024contrastive}. For all methods,} in line with previous works~\citep{barbano2023contrastive,trauble2024contrastive}, we use the Adam optimizer, with an initial learning rate of $10^{-4}$ decayed by a factor of $0.9$ every 10 epochs, and a weight decay of $5\times 10^{-5}$. We use a batch size of 32, and train for a total of 300 epochs. \rebuttal{For contrastive losses we employ a temperature of 2.0 for RNC~\citep{zha2023rank} and 0.1 for all others.}
Our experiments are implemented in PyTorch and run on the Leonardo and Jean Zay clusters on A100 and V100 NVIDIA GPUs. %

\paragraph{Evaluation metrics}
We evaluate the mean absolute error (MAE), the $R^2$ coefficient \rebuttal{pearson correlation $r$} on the test set of each dataset. Furthermore, the OpenBHB challenge computes the MAE on both the internal (Int. MAE) and external (Ext. MAE) test sets and the balanced accuracy (BAcc) on site prediction, training a logistic regression on the model representations. %
The final challenge score is then computed as $\mathcal{L}_c = \text{BAcc}^{0.3} \cdot \text{MAE}_{\text{ext}}$. On AD and MCI patients, we also consider the ROC curves computed on BAG, and fine-tuning accuracy for classification. %

\begin{table*}
    \centering
    \caption{\textbf{Brain age regression results.} 
    Single means that the models were trained and tested on each dataset separately; \emph{Merge} means that the models were trained on the combined healthy samples.
    OpenBHB Challenge Results report balanced accuracy (BAcc), internal MAE, external MAE, and the composite score $\mathcal{L}_{c}$. \rebuttal{For single training, the challenge evaluation was performed on models trained on OpenBHB.}}
    \label{tab:regression-results-summary}
    \resizebox{\linewidth}{!}{%
    \begin{tabular}{l c ccc ccc ccc ccc | c c c c}
    \toprule
    & & \multicolumn{3}{c}{OpenBHB} & \multicolumn{3}{c}{ADNI} & \multicolumn{3}{c}{ICBM} & \multicolumn{3}{c}{OASIS3} & \multicolumn{4}{c}{\textbf{OpenBHB Challenge Results}} \\
    Method & Train & MAE & $R^2$ & \rebuttal{r (corr.)} & MAE & $R^2$ & \rebuttal{r (corr.)} & MAE & $R^2$ & \rebuttal{r (corr.)} &MAE & $R^2$ & \rebuttal{r (corr.)} & \textbf{BAcc} & \textbf{Int. MAE} & \textbf{Ext. MAE} & $\mathcal{L}_{c}\,(\downarrow)$ \\
    \midrule
    L1 & \multirow{6}{*}{Single}
        & 2.73 & 0.92 & \rebuttal{0.96} & 3.64 & 0.58 & \rebuttal{0.78} & 4.33 & 0.75 & \rebuttal{0.87} & 4.32 & 0.65 & \rebuttal{0.82}
        & 6.7 & 2.67 & 4.18 & 1.82 \\
    $\mathcal{L}^{y-aware}$~\citep{dufumier2021contrastive}
        & & \underline{2.65} & 0.92 & \rebuttal{0.96} & 3.47 & 0.61 & \rebuttal{0.79} & \underline{4.21} & 0.76 & \rebuttal{0.87} & 4.26 & 0.65 & \rebuttal{0.82}
        & 6.60 & {2.66} & \underline{4.10} & 1.82 \\
    $\mathcal{L}^{threshold}$~\citep{barbano2023contrastive}
        & & {2.66} & 0.93 & \rebuttal{0.96} & \textbf{3.36} & 0.63 & \rebuttal{0.65} & 5.27 & 0.57 & \rebuttal{0.86} & \textbf{4.18} & 0.66 & \rebuttal{0.82}
        & 5.73 & 2.95 & \underline{4.10} & 1.74 \\ 
    $\mathcal{L}^{exp}$~\citep{barbano2023contrastive} 
        & & {2.66} & 0.93 & \rebuttal{0.96} & \underline{3.44} & 0.62 & \rebuttal{0.79} & \textbf{4.11} & 0.77 & \rebuttal{0.88} & \underline{4.22} & 0.66 & \rebuttal{0.82}
        & \underline{5.1} & \textbf{2.55} & \textbf{3.76} & \textbf{1.54} \\
    RNC~\citep{zha2023rank} 
        & & \rebuttal{3.78} & \rebuttal{0.86} & \rebuttal{0.56} & \rebuttal{3.73} & \rebuttal{0.56} & \rebuttal{0.45} & \rebuttal{4.57} & \rebuttal{0.72} & \rebuttal{0.51} & \rebuttal{4.48} & \rebuttal{0.63} & \rebuttal{0.49} & \rebuttal{\textbf{4.9}} & \rebuttal{3.63} & \rebuttal{4.30} & \rebuttal{1.74} \\
    $\mathcal{L}^{AdapN}$~\citep{trauble2024contrastive} 
        & & \rebuttal{\textbf{2.54}} & \rebuttal{0.93} & \rebuttal{0.58} & \rebuttal{3.47} & \rebuttal{0.61} & \rebuttal{0.79} & \rebuttal{4.59} & \rebuttal{0.71} & \rebuttal{0.86} & \rebuttal{4.30} & \rebuttal{0.65} & \rebuttal{0.82} & \rebuttal{5.53} & \rebuttal{\underline{2.61}} & \rebuttal{4.16} & \rebuttal{1.75} \\
    \midrule\midrule
    L1 & \multirow{6}{*}{Merge}
        & \textbf{2.59} & 0.93 & \rebuttal{0.97} & 3.45 & 0.61 & \rebuttal{0.79} & 3.77 & 0.82 & \rebuttal{0.91} & 4.08 & 0.68 & \rebuttal{0.84}
        & 9.33 & \textbf{2.42} & 2.43 & 1.19 \\
    $\mathcal{L}^{y-aware}$~\citep{dufumier2021contrastive}
        & & \underline{2.70} & 0.91 & \rebuttal{0.96} & \underline{3.40} & 0.63 & \rebuttal{0.80} & \underline{3.73} & 0.82 & \rebuttal{0.92} & 4.08 & 0.67 & \rebuttal{0.84}
        & 7.3 & {2.61} & 2.53 & 1.15 \\
    $\mathcal{L}^{threshold}$~\citep{barbano2023contrastive}
        & & 3.11 & 0.85 & \rebuttal{0.94} & 3.43 & 0.62 & \rebuttal{0.80} & 3.83 & 0.80 & \rebuttal{0.91} & 4.11 & 0.66 & \rebuttal{0.84}
        & 6.0 & 2.90 & \textbf{2.21} & 0.95 \\ 
    $\mathcal{L}^{exp}$~\citep{barbano2023contrastive}
        & & {2.77} & 0.91 & \rebuttal{0.95} & \textbf{3.39} & 0.63 & \rebuttal{0.80} & 3.84 & 0.81 & \rebuttal{0.91} & \underline{4.05} & 0.68 & \rebuttal{0.85}
        & \underline{5.40} & 2.71 & \underline{2.25} & \textbf{0.93} \\
    RNC~\citep{zha2023rank} 
        & & \rebuttal{4.92} & \rebuttal{0.77} & \rebuttal{0.70} & \rebuttal{5.57} & \rebuttal{-0.01} & \rebuttal{0.42} & \rebuttal{4.95} & \rebuttal{0.70} & \rebuttal{0.67} & \rebuttal{6.39} & \rebuttal{0.19} & \rebuttal{0.38} & \rebuttal{\textbf{2.7}} & \rebuttal{5.08} & \rebuttal{6.13} & \rebuttal{2.07} \\
    $\mathcal{L}^{AdapN}$~\citep{trauble2024contrastive} 
        & & \rebuttal{2.71} & \rebuttal{0.92} & \rebuttal{0.96} & \rebuttal{3.42} & \rebuttal{0.62} & \rebuttal{0.80} & \rebuttal{\textbf{3.71}} & \rebuttal{0.83} & \rebuttal{0.92} & \rebuttal{\textbf{3.96}} & \rebuttal{0.69} & \rebuttal{0.84} & \rebuttal{6.37} & \rebuttal{\underline{2.54}} & \rebuttal{2.36} & \rebuttal{1.03} \\
    \bottomrule
    \end{tabular}
    }
\end{table*}

\subsection{Proposed method}
\label{sec:method}

\rebuttal{In this section, we present our contrastive learning framework for regression, originally introduced in~\cite{barbano2023contrastive}, leading to the formulation of our proposed loss function $\mathcal{L}^{exp}$.}

\paragraph{Background}
\rebuttal{In contrastive learning, given a collection of samples (here, images) $x \in \mathcal{X}$, the objective is to learn a parametric function $f: \mathcal{X} \rightarrow \mathbb{S}^{d-1}$ that projects similar, or \textit{positive}, samples $x^+$ close together in a shared embedding space (the hypersphere $\mathbb{S}^{d-1}$), while pushing \textit{negative} samples $x^-$ apart~\citep{khosla2020supervised,barbano2023unbiased}. %
For a given anchor sample $x_i$ with latent representation $f(x_i)$, the model encourages high similarity with embeddings of samples sharing the same label and low similarity with those from other categories. 
This formulation has proven highly effective for classification problems; however, it is not directly applicable to regression, where labels are \textit{continuous} (i.e., $y \in \mathbb{R}$) and the distinction between positive and negative pairs is inherently ambiguous.}
\rebuttal{To solve this issue, in~\cite{barbano2023contrastive} we propose to use} a \emph{degree of similarity} defined as a kernel function $K$ on the continuous labels $y$:
\begin{equation}
w_k = K(y - y_k), \quad \text{where } 0 \leq w_k \leq 1
\end{equation}
\noindent Here, \( w_k \) encodes how similar the sample \( x_k \) is to the anchor \( x_i \) in label space. As for $K$, we choose a Radial Basis Function (RBF) parametrized by $\sigma$, chosen by the user, but other choices can be made. Our goal becomes to learn a mapping \( f \) such that samples with high degree of similarity \( w_k \) are embedded close to the anchor in the representation space $\mathbb{S}^{d-1}$, while those with low \( w_k \) are pushed farther apart.

\paragraph{Loss Formulations}
\rebuttal{From this idea, we define different loss functions.}
We begin from the general contrastive condition~\citep{barbano2023unbiased}:
\begin{equation}
    s^-_t - s^+_k \leq 0 \quad \forall t,k
    \label{eq:eps-condition}
\end{equation}
where \( s_k = \text{sim}(f(x), f(x_k)) \) is the similarity between the anchor \( x_i \) and a sample \( x_k \), $s^-_t$ denotes the similarity between $x_i$ and the negative $x^-_t$, and $s^+_k$ with the positive $x^+_k$. Eq.~\eqref{eq:eps-condition} models the main idea of contrastive learning, where the similarity between the embeddings of positive samples should be bigger than negative ones. Eq.~\eqref{eq:eps-condition} can be transformed in an optimization problem using the $\max$ operator and its smooth approximation \textit{LogSumExp} (details can be found in~\cite{barbano2023unbiased}) resulting in the well-known InfoNCE loss~\citep{chen2020simple}: 
\begin{equation}
    \argmin_f \sum_k \max (0,\{  s^-_t - s^+_k\}_{\forall t})  \approx -\sum_k \log \left(\frac{\exp s^+_k}{\exp s^+_k + \sum_{t}\exp s^-_t} \right)
\end{equation}
This formulation is well suited for classification but not for regression, since one would need well separated classes (thus discrete labels). %

\paragraph{\( \mathcal{L}^{y\text{-aware}} \)~\citep{dufumier2021contrastive}}
A straightforward idea to tackle regression with continuous labels is to softly weight \rebuttal{alignment of samples} by their similarity to the anchor:
\begin{equation}
\begin{aligned}
    \mathcal{L}^{y\text{-aware}} &= - \sum_k \frac{w_k}{\sum_t w_t}
    \log \left( \frac{\exp(s_k)}{\sum_{t=1}^N \exp(s_t)} \right)
\end{aligned}
\end{equation}

This \rebuttal{loss} generalizes contrastive learning to continuous attributes. However, its denominator \rebuttal{uniformly repels samples apart from the anchor, disregarding their similarity, as measured by the weights $(w_t)_{t\neq k}$. This might be suboptimal for regression.}

\paragraph{\rebuttal{Non-uniform hard thresholding}}
\rebuttal{To address this, we propose to include in the denominator only samples with lower degrees of similarity than $x_k$, i.e., with \( w_t < w_k \)}:
\begin{equation}
\begin{aligned}
    \mathcal{L}^{thresold} = -\sum_k \frac{w_k}{\sum_t \delta_{w_t < w_k} w_t}
    \log \left( \frac{\exp(s_k)}{\sum_{t \neq k} \delta_{w_t < w_k} \exp(s_t)} \right)
\end{aligned}
\end{equation}
This prevents repelling samples more similar than \( x_k \). \rebuttal{Nonetheless, because of the exponential in the denominator, it sill emphasizes the closest negative just under the threshold, which is not ideal.}

\paragraph{\rebuttal{Non-uniform soft thresholding}}
\rebuttal{In~\cite{barbano2023contrastive}}, we introduce an exponential formulation that modulates repulsion strength based on dissimilarity:
\begin{equation}
\begin{aligned}
    \mathcal{L}^{exp} = -\frac{1}{\sum_t w_t} \sum_k w_k \log
    \left( \frac{\exp(s_k)}{\sum_{t \neq k} \exp(s_t(1 - w_t))} \right)
\end{aligned}
\end{equation}
Here, the factor \( (1 - w_t) \) acts as a soft inverse similarity, increasing the repulsion for dissimilar samples and attenuating it for similar ones. This formulation avoids hard thresholds and ensures that nearby samples are preserved in the representation space, making it particularly well-suited for regression tasks.

\section{Results}

\begin{figure*}
\begin{subfigure}{0.49\linewidth}
    \centering
    \includegraphics[width=\linewidth]{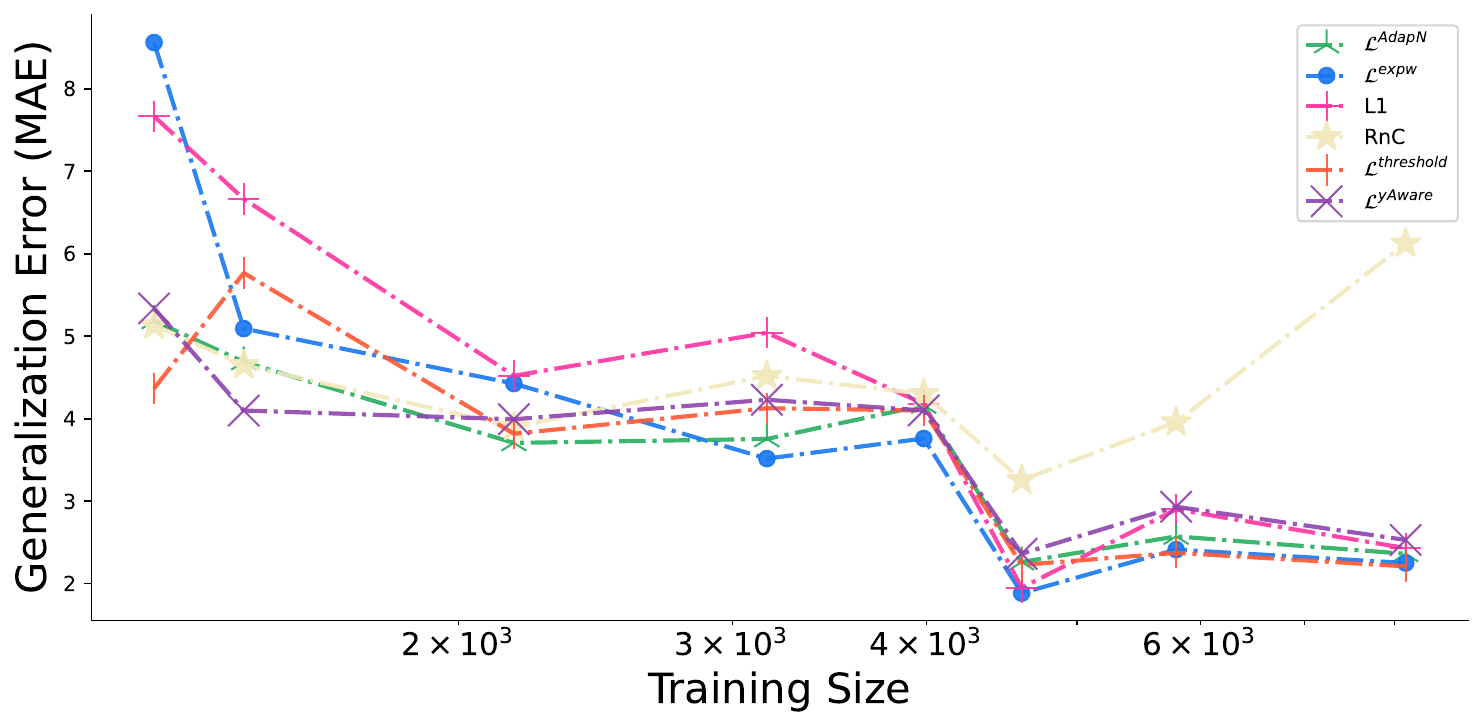}
    \caption{External MAE vs training size (scaling)}
    \label{fig:mae-scaling}
\end{subfigure}
\begin{subfigure}{0.49\linewidth}
    \centering
    \includegraphics[width=\linewidth]{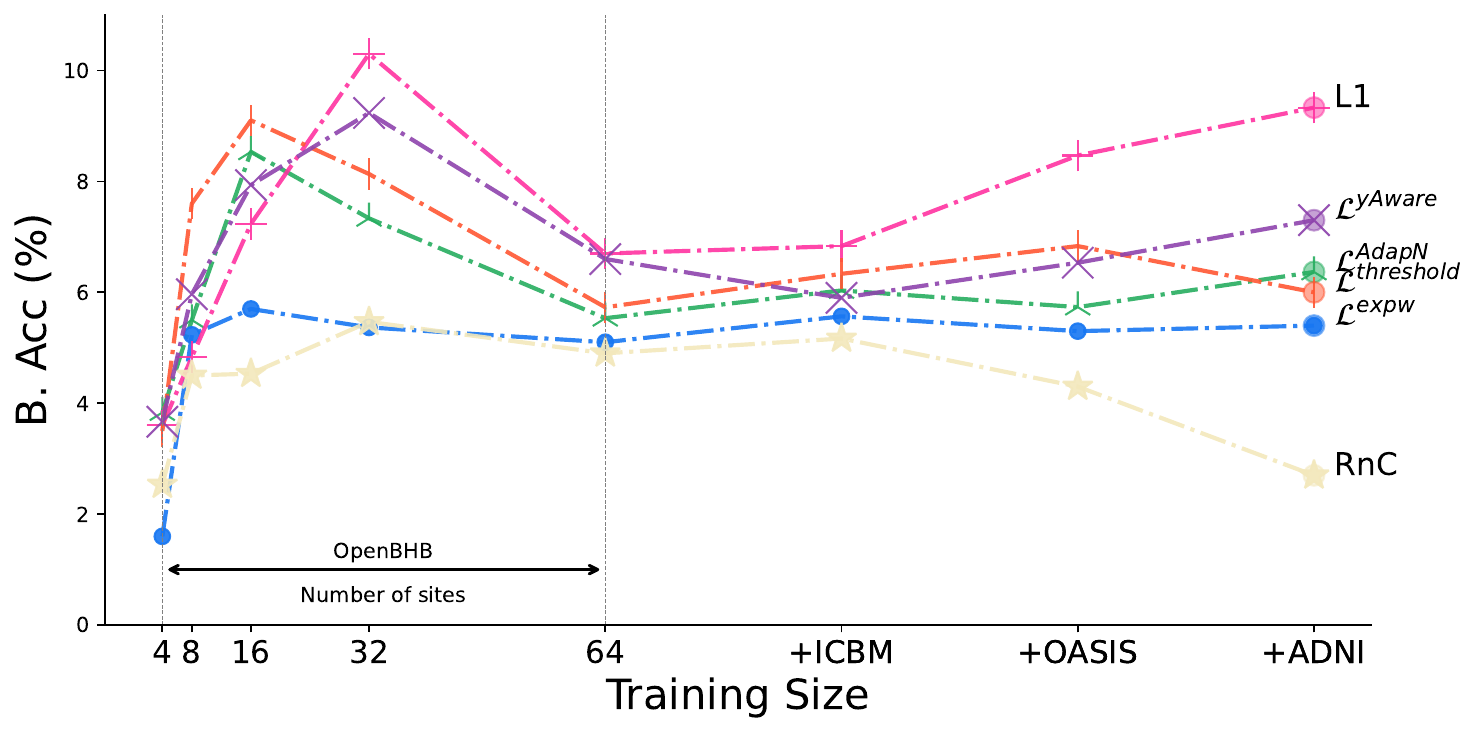}
    \caption{Site bias vs training size. $\mathcal{L}^{exp}$ stays consistent.}
    \label{fig:bacc-scaling}
\end{subfigure}
\caption{\rebuttal{\textbf{Scaling contrastive brain-age models with data size.}
\emph{(a)} External mean absolute error (MAE) decreases as training size increases; all methods besides RnC benefit from more data, reaching a MAE between 2.21 and 2.53. 
\emph{(b)} Balanced accuracy (B. Acc) in predicting scanner site increases for most methods as training size grows, indicating increased site bias. 
In contrast, \(\mathcal{L}^{exp}\) remains stable and low across all sizes, demonstrating robustness to site-specific confounds. On the other hand, the decrease in B. Acc for RnC can be attributed to the degraded performance in brain age prediction when datasets are merged (e.g., the model collapsed and did not learn meaningful representations). In fact RnC shows a trend similar to other methods when trained on OpenBHB only. We hypothesize this is due to the unweighted alignment term and to the high temperature value used by RnC, which could focus the repulsion on samples coming from different dstributions (e.g., ADNI and OASIS subjects are older), leading to a latent space which does not capture well biological and age variability.}}
\label{fig:scaling}
\end{figure*}

\begin{figure*}
    \begin{subfigure}{\linewidth}
        \includegraphics[width=0.24\linewidth]{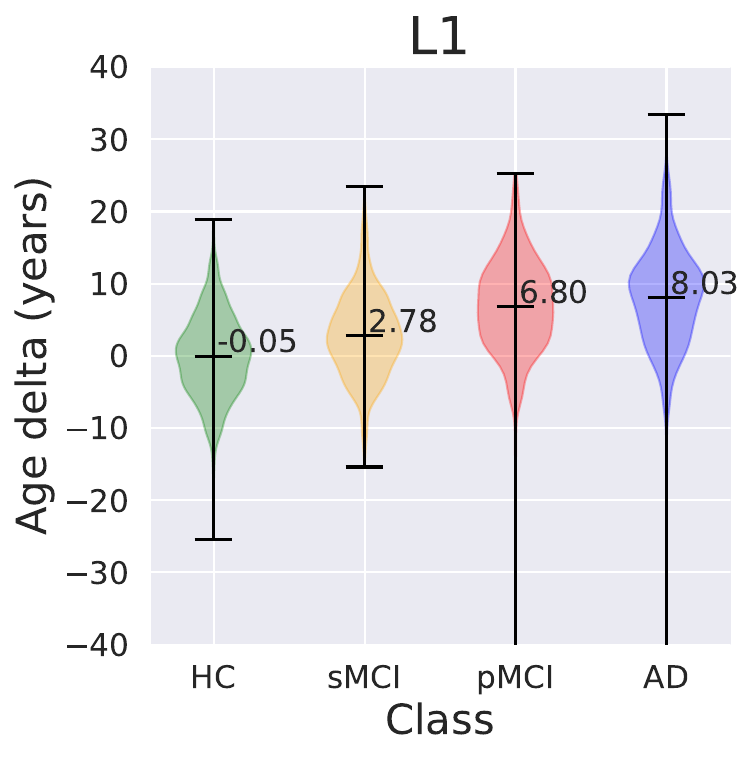}
        \includegraphics[width=0.24\linewidth]{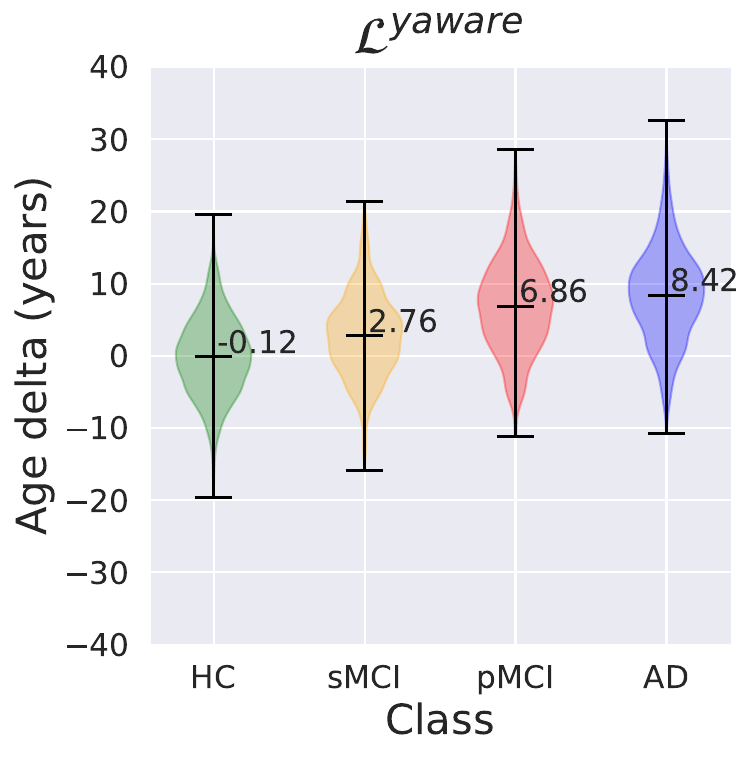}
        \includegraphics[width=0.24\linewidth]{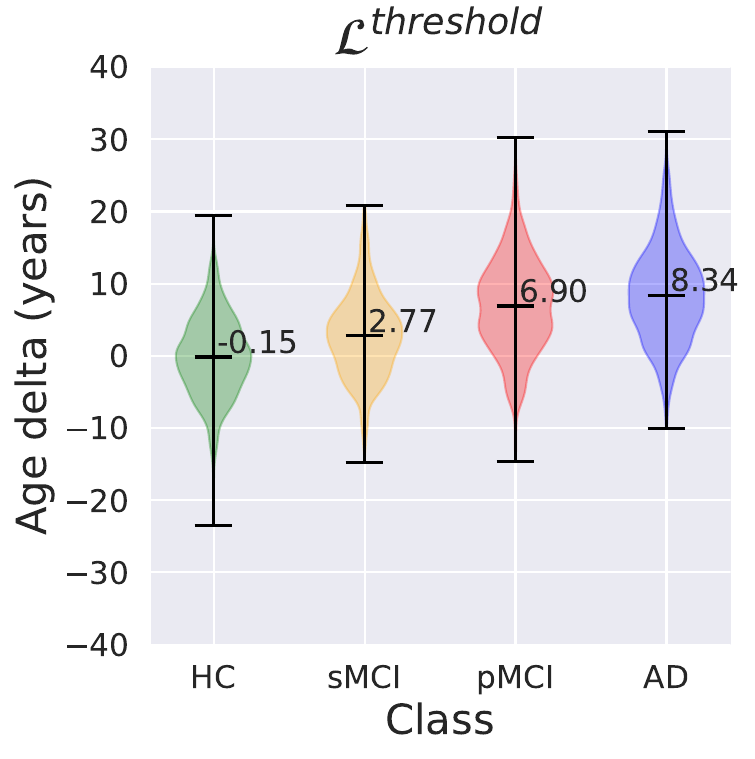}
        \includegraphics[width=0.24\linewidth]{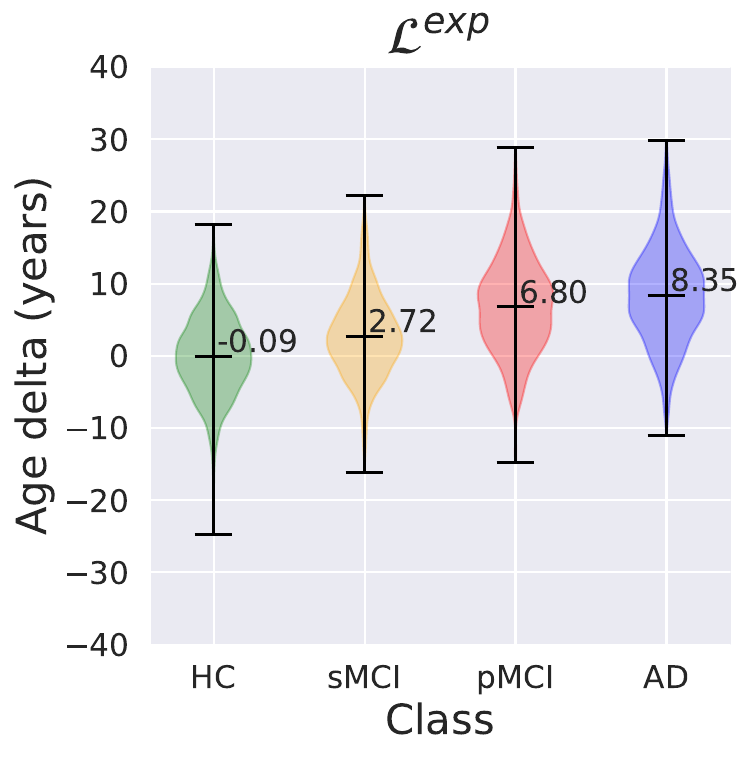}
        \caption{\textbf{\rebuttalrev{Brain Age Gap} (BAG) distribution for HC, MCI, and AD classes across all datasets.} The shaded areas represent the age delta distribution (histogram) for each class. The horizontal black line in the middle highlights the mean value (explicitly annotated for readability).}
        \label{fig:class-mean-age-delta}
    \end{subfigure}
    
    \vspace{1em}
    
    \begin{subfigure}[b]{0.5\linewidth}
        \includegraphics[width=0.48\linewidth]{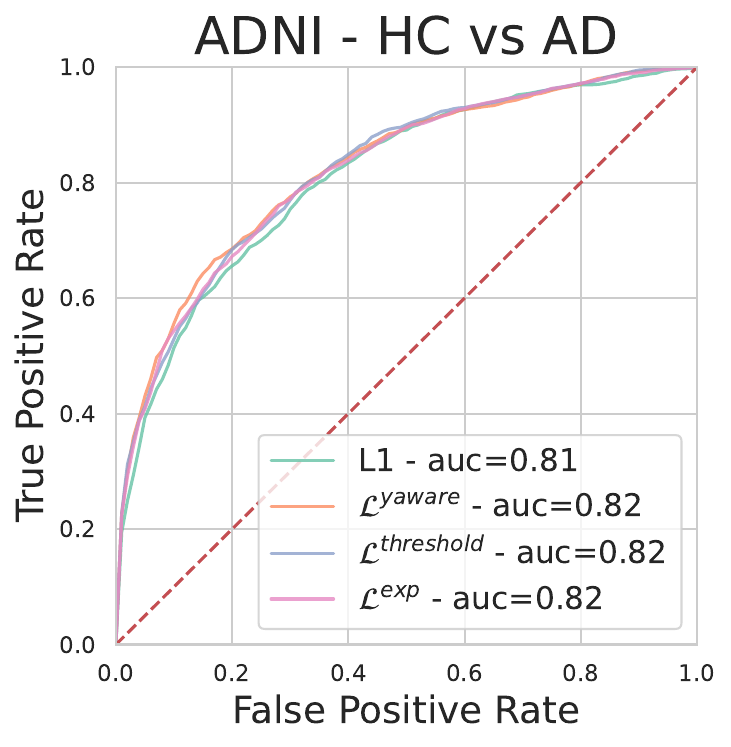}
        \includegraphics[width=0.48\linewidth]{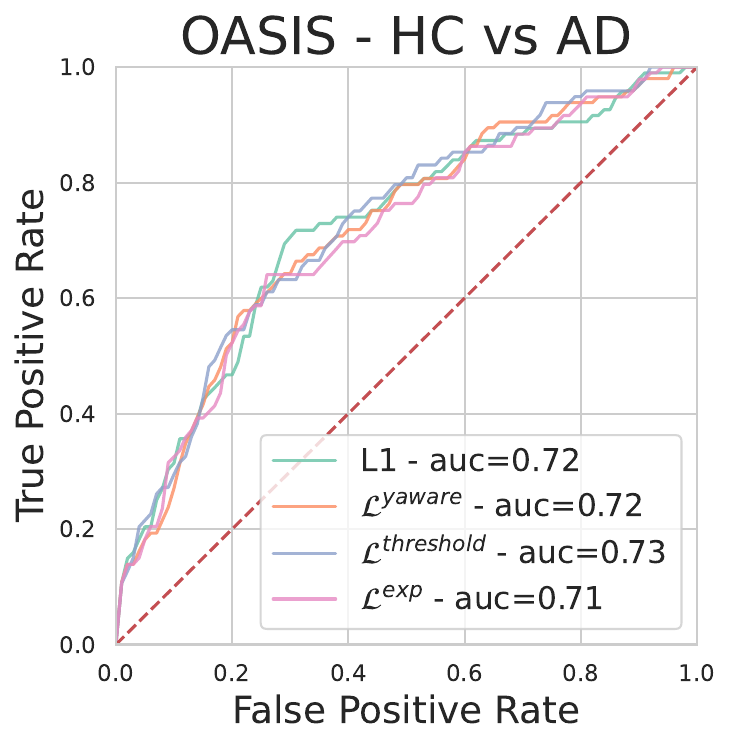}
        \caption{ROC plots for using BAG as a proxy for detecting AD patients.}
    \label{fig:bag-rocauc-bigmri}
    \end{subfigure}
    \hfill
    \begin{subfigure}[b]{0.48\linewidth}
        \includegraphics[width=\linewidth]{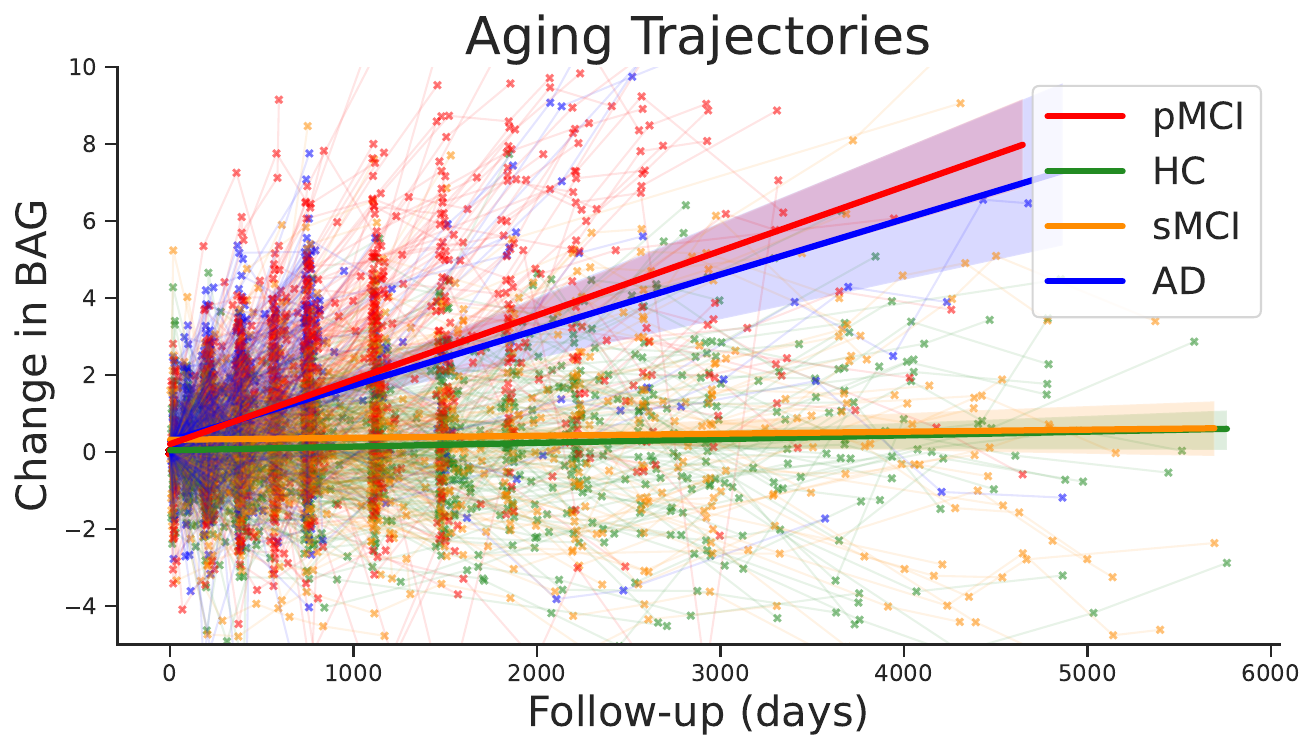}
        \caption{Longitudinal change in BAG on ADNI. We only considered patients with a minimum of 3 follow-up exams. %
        }
        \label{fig:bag-adni-longitudinal}
    \end{subfigure}
        \caption{
\textbf{Using \rebuttalrev{BAG} as a proxy for detecting neurodegeneration.} 
(a) BAG distribution for HC, stable MCI (sMCI), progressive MCI (pMCI), and AD patients across all datasets. All methods consistently show increased BAG in pMCI and AD groups, indicating accelerated brain aging. 
(b) ROC curves evaluating the ability of BAG to separate HC and AD subjects in ADNI and OASIS datasets. Contrastive methods (\( \mathcal{L}^{exp}, \mathcal{L}^{threshold}, \mathcal{L}^{yaware} \)) match or slightly outperform the L1 baseline. 
(c) Longitudinal BAG trajectories (with $\mathcal{L}^{exp})$ on ADNI, showing that pMCI and AD subjects exhibit increased BAG over time, while HC and sMCI groups remain stable. This confirms that contrastive models capture disease progression dynamics.
}
    \label{fig:bag-adni}
\end{figure*}

We base our analysis on a collection of different publicly available datasets (OpenBHB, ADNI, ICBM, OASIS) for a total of 7,045 subjects and 20,647 structural MRI scans.

\subsection{Scaling training size improves generalization} 
In Tab.~\ref{tab:regression-results-summary} we report results of brain age regression across the different datasets employed. We experiment with two pre-training schemes, namely, we first train and test on the same dataset (single), and then we train on the combined healthy population across all datasets and test on each test set separately (merge). As we can note from the results, the overall error decreases when pretraining on the larger dataset. We also benchmark our models on the public OpenBHB challenge, where we significantly improve the results in terms of generalization (external MAE). With $\mathcal{L}^{exp}$, we achieve state-of-the-art results in the challenge, scoring an impressive 0.93, compared to the previous SOTA of 1.54. By increasing the training set size, we are able to almost halve the external prediction error, meaning that the models generalize better. Fig.~\ref{fig:mae-scaling} shows models scaling in terms of external MAE compared to the training set size. As we can see, almost all methods benefit from increasing the pre-training size. This is consistent with related literature~\citep{schulz2024performance}, showing that deep models follow exponential laws in brain age prediction from T1 structural MRIs. \rebuttal{This does always hold for RnC, which collapses when datasets are merged. We hypothesize this is due to the unweighted alignment term and to the high temperature value~\citep{zha2023rank}.}

\subsection{$\mathcal{L}^{exp}$ is more invariant to site bias}

By closely inspecting results in Tab.~\ref{tab:regression-results-summary}, we notice an interesting phenomenon: all models pre-trained on the merged dataset, other than $\mathcal{L}^{exp}$, exhibit a higher balanced accuracy (Bacc) in retrieving the acquisition site of the scan, meaning that as the training size gets larger, they become more biased towards the site. $\mathcal{L}^{exp}$ is the only method which preserve a final balanced accuracy below 6\%. To illustrate these results in more detail, in Fig.~\ref{fig:bacc-scaling}, we show how the final balanced accuracy scales with respect to the size of the training data, in terms of acquisition sites and datasets. As we can observe from the results, our proposed loss $\mathcal{L}^{exp}$ shows a balanced accuracy almost constant at different training size, whereas most other methods are heavily influenced by the training composition, and tend to become more biased as diversity in the data increases. 
We hypothesize that this is due to the difference in the formulation of $\mathcal{L}^{exp}$ with respect to the other losses, such as y-aware. The presence of the exponential in the uniformity term of y-aware~\citep{dufumier2021contrastive} may force some samples to be pushed apart from the anchor more than they should. This could result in a less ``compact'' latent space, that might capture more variance in the input data, such as noise from the site effect. On the other hand, $\mathcal{L}^{exp}$ may avoid this phenomenon as the samples are repelled according to the kernel value~\citep{barbano2023contrastive}, achieving latent representations that are more invariant to such noise.

\begin{table}
    \centering
    \caption{\rebuttal{\textbf{Downstream results (AD detection) and cross-dataset generalization} in terms of balanced accuracy (\%). Models are pre-trained on OpenBHB, finetuned on ADNI or OASIS, and tested on each dataset. Results show that CL methods are always competitive or better than L1 pre-training.}}
    \resizebox{\linewidth}{!}{%
    \begin{tabular}{lccccc}
    \toprule
     & ADNI $\rightarrow$ ADNI & ADNI $\rightarrow$ OASIS & OASIS $\rightarrow$ ADNI & OASIS $\rightarrow$ OASIS & Avg. \\
    \midrule
    L1 & \textbf{86.8}\% \std{2.8\%} & 50.0\% \std{0.0\%} & 53.5\% \std{11.3\%} & 62.0\% \std{4.4\%} & 63.1\% \\
    $\mathcal{L}^{yAware}$ & 85.0\% \std{2.5\%} & \underline{60.7}\% \std{7.2\%} & 71.1\% \std{13.5\%} & 67.0\% \std{5.8\%} & 70.9\% \\
    $\mathcal{L}^{threshold}$ & 84.9\% \std{2.4\%} & 51.9\% \std{1.9\%} & 65.1\% \std{8.9\%} & \underline{67.8}\% \std{7.8\%} & 67.4\% \\
    $\mathcal{L}^{expw}$ & 85.7\% \std{3.0\%} & \textbf{68.8}\% \std{6.7\%} & 66.1\% \std{12.5\%} & \textbf{71.8}\% \std{5.0\%} & \textbf{73.1}\% \\
    $\mathcal{L}^{AdapN}$ & 85.4\% \std{2.4\%} & 51.4\% \std{1.6\%} & \underline{82.8}\% \std{2.4\%} & 62.7\% \std{4.9\%} & 70.6\% \\
    RnC & \underline{85.8}\% \std{2.6\%} & 55.1\% \std{4.1\%} & \textbf{83.5}\% \std{1.6\%} & 64.2\% \std{5.1\%} & \underline{72.1}\% \\
    \bottomrule
    \end{tabular}
    }
    \label{tab:ad-cross-dataset}
\end{table}

\begin{figure}
    \centering
    \includegraphics[width=\linewidth]{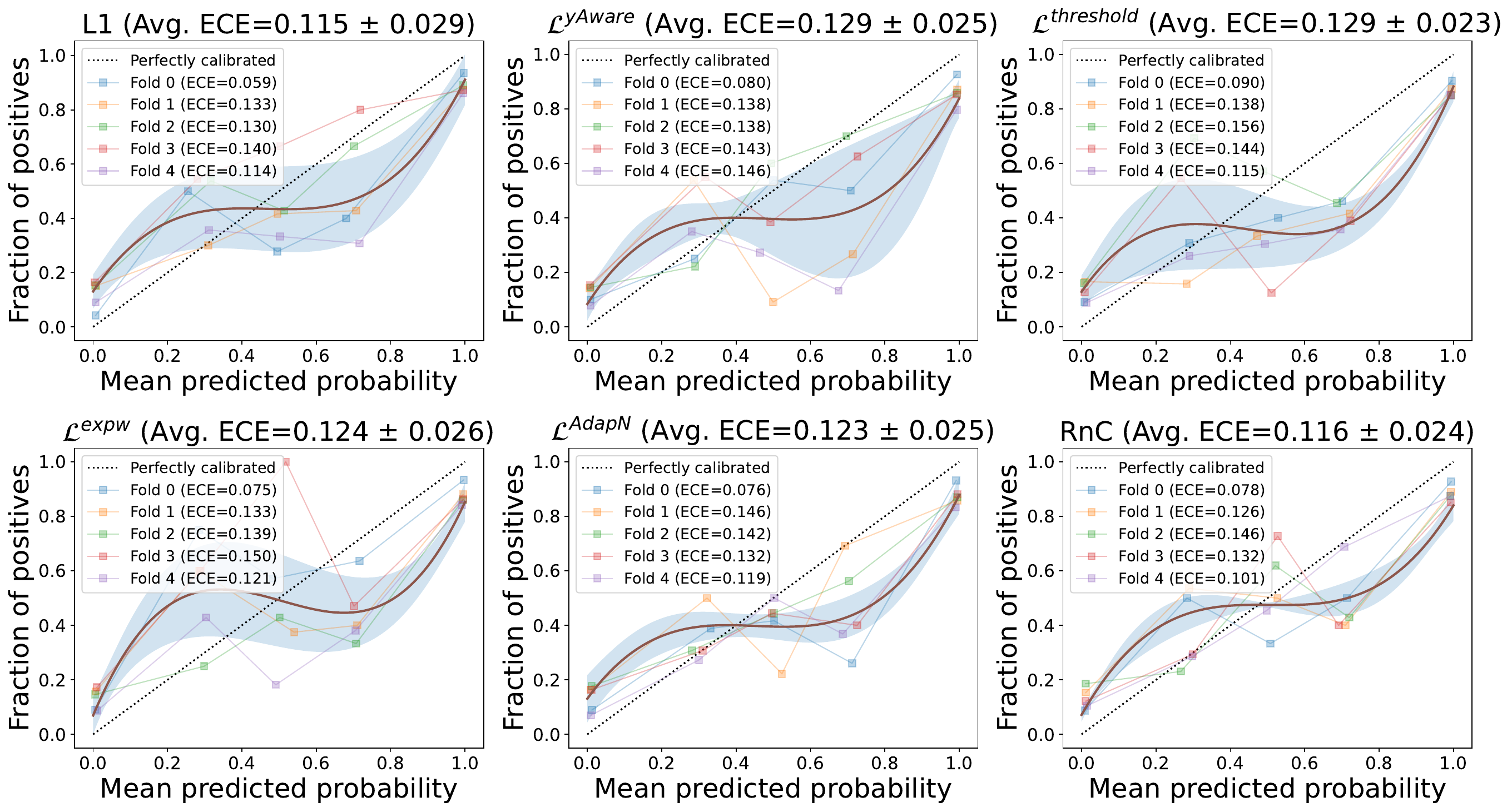}
    \caption{\rebuttal{\textbf{Calibration curves} show that all fine-tuned models are reasonably calibrated on the downstream task, with an Average Calibration Error (ECE) slightly higher than 10\%.}}
    \label{fig:calibration}
\end{figure}
\begin{figure}
    \centering
    \includegraphics[width=\linewidth]{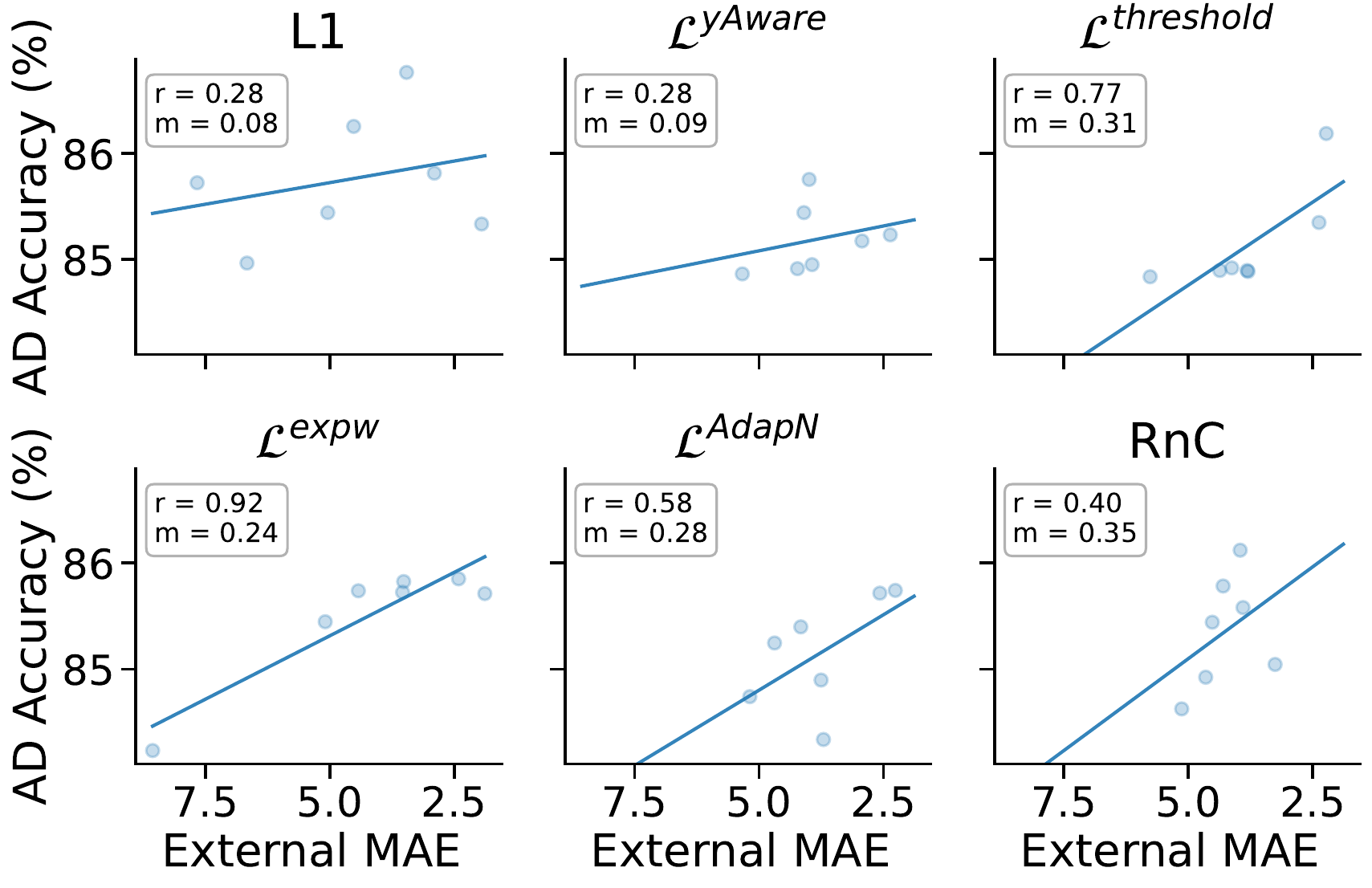}
    \caption{\rebuttal{\textbf{Correlation} between downstream accuracy (ADNI) and brain age prediction error. Contrastive methods such as $\mathcal{L}^{exp}$ benefit from lower MAE.}}
    \label{fig:acc-vs-mae}
\end{figure}

\subsection{CL is a reliable estimator of accelerated aging}
Next, we aim to assess whether CL is a reliable estimator to detect accelerated aging in subjects affected by cognitive impairment and neurodegenerative diseases such as Alzheimer's Disease. Brain age is widely recognized as a biomarker of different brain disorders~\citep{kaufmann2019common}, and many works proposed deep learning pipelines for brain age estimation~\citep{cole2017predicting, he2021global}. These models are typically trained with a supervised loss such as L1 and have shown robust performance in highlighting accelerated aging in affected patients. To assess whether our proposed CL framework also exhibits this property, we employ data from the Alzheimer’s Disease Neuroimaging Initiative (ADNI) with 1754 subjects and from OASIS-3 with 685 subjects for a total of 2439 subjects. The analysis is presented in Fig.~\ref{fig:bag-adni}. First, in Fig.~\ref{fig:class-mean-age-delta}, we show the distribution of brain age gap (BAG) - the difference between the model's predicted age and the subject's chronological age - across all groups HC, stable mild cognitive impairment (sMCI), progressive mild cognitive impairment (pMCI), and AD. The obtained results are mostly consistent among all methods, showing that contrastive losses are comparable to standard L1 supervised models. Furthermore, we notice that L1 exhibits a higher spread on the pMCI and AD classes, compared to CL methods. In Fig.~\ref{fig:bag-rocauc-bigmri} we report the ROC curves for detecting AD patients on ADNI and OASIS. Once more, the results show that contrastive learning methods are comparable to standard L1 supervised baselines, even with a slight edge (AUC of 0.82 vs 0.81 on ADNI and 0.73 vs 0.72 on OASIS). Lastly, we assess the longitudinal reliability of $\mathcal{L}^{exp}$ by calculating the change in BAG during follow-ups for ADNI subjects. As expected from~\cite{franke2012longitudinal}, we observe an increase of BAG for AD and pMCI patients, meaning an increased brain age during follow-ups; whereas HC and sMCI subjects remain stable.

\subsection{Do better brain age models translate to better diagnostic models?}

Finally, we also analyze whether more accurate brain age prediction models can translate to better diagnostic models. To do so, we finetune the models pre-trained on OpenBHB using ADNI \rebuttal{and OASIS data. Results are reported in Tab.~\ref{tab:ad-cross-dataset}. Results show that CL is almost always better than L1 pre-training. The proposed $\mathcal{L}^{expw}$ achieves the best generalization in two out of four transfer directions, reaching accuracies of 68.8\% and 71.8\%, respectively, and achieves the highes average accuracy. Similarly, RnC and $\mathcal{L}^{AdapN}$ perform competitively when transferring from OASIS to ADNI, with accuracies exceeding 82\%. On the other hand, L1 achieves results closer to a random guess in both cross-dataset generalization tasks. These results confirm that contrastive pre-training yields more transferable representations that maintain discriminative power across acquisition sites and cohorts.
In terms of model calibration (Fig.~\ref{fig:calibration}), all methods remain reasonably well-calibrated, with Expected Calibration Error (ECE) values around 10–12\% on ADNI.} We also report the downstream accuracy in classifying AD and HC versus the original brain age MAE in Fig.~\ref{fig:acc-vs-mae}. 
Consistently with related literature~\citep{bashyam2020mri}, we find that standard L1 supervised models do not really benefit on average from lower age prediction error, showing a correlation between age MAE and classification accuracy of $r=0.28$. In fact, \cite{bashyam2020mri,schulz2024beyond} note that tighter fitting models on age naturally produce estimates that offer little information beyond age, and moderately fitting brain age models obtain significantly higher differentiation in a transfer-learning setup. On the contrary, our results seem to suggest that models pre-trained using CL can learn complex and meaningful representations that can benefit downstream tasks. Specifically, $\mathcal{L}^{exp}$ shows an almost total correlation of $r=0.92$ between age prediction error and downstream accuracy, and almost all CL methods exhibit a significantly higher slope than the L1 baseline (meaning that the increase in accuracy is more rapid). This shows that CL is less prone to overfitting the brain age prediction task. This is an important finding, as it may lay the foundation for robust pre-trained pipelines on neuroimaging data.

\section{Discussion}

Our study focused on how CL can be scaled to deliver state-of-the-art performance on brain-age regression, while also producing embeddings that remain useful for downstream clinical applications. Four broad findings emerge.

\paragraph{Large, heterogeneous pre-training data consistently improve brain-age estimation and downstream generalization}
\rebuttalrev{On all benchmarks}, training on the merged healthy cohort reduced the error relative to the single-dataset baseline (Tab.~\ref{tab:regression-results-summary}).
The gains are most pronounced on the OpenBHB challenge, where the external MAE nearly halves on average.
This mirrors the scaling laws recently reported for \rebuttalrev{L1-}supervised models \citep{schulz2024performance}, and shows that CL obeys similar dynamics.
Importantly, Fig.~\ref{fig:mae-scaling} shows that performance does not saturate in the range we explored, suggesting further room for improvement as larger neuroimaging datasets become available.

\paragraph{$\mathcal{L}^{exp}$ mitigates site bias at scale}
While every loss benefited from more data, only $\mathcal{L}^{exp}$ maintained a low balanced accuracy ($<6\%$) in predicting acquisition site after pre-training (Tab.~\ref{tab:regression-results-summary}, Fig.~\ref{fig:bacc-scaling}).
We hypothesize that the weighted repulsion term in $\mathcal{L}^{exp}$ prevents over-dispersion of negatives, yielding a latent space that captures biologically relevant variance while suppressing site-specific factors.
By contrast, the fixed uniformity term in $y$-Aware and $\mathcal{L}^{threshold}$ can force hard negatives to drift apart excessively, leaving room for confounding factors to be learned. In this regard, L1 shows the strongest bias towards site.
In practice, this robustness is crucial when models are deployed across centres that differ in hardware and protocol.

\paragraph{\rebuttalrev{CL} pre-training is a reliable detector of accelerated ageing}
All \rebuttalrev{CL} models replicated the well-established pattern of higher brain-age gaps in pMCI and AD groups (Fig.~\ref{fig:class-mean-age-delta}), also showing slightly lower variance than the L1 baseline.
BAG-based ROC curves (Fig.~\ref{fig:bag-rocauc-bigmri}) equal or surpass the L1 reference (AUC 0.82 vs. 0.81 on ADNI; 0.73 vs. 0.72 on OASIS-3).
Moreover, longitudinal trajectories on ADNI confirm that $\mathcal{L}^{exp}$ captures progressive neurodegeneration (Fig.~\ref{fig:bag-adni-longitudinal}).
Taken together, these results indicate that \rebuttalrev{CL} embeddings preserve clinically meaningful signals rather than merely fitting chronological age.

\paragraph{Lower brain-age MAE translates into higher diagnostic accuracy, when the model is contrastive}
For L1 models, tighter age fits do not necesserily predict better HC/AD separation, consistent with prior reports that over-fitting to age harms downstream transfer \citep{bashyam2020mri, tan2025mind}.
Interestingly, the opposite seems to hold for $\mathcal{L}^{exp}$ ($r=0.92$).
We argue that CL pre-training encourages more \rebuttalrev{general} representations, whereas strongly supervised objectives push all capacity toward age alone.
This finding positions CL models as promising foundation models for a broad range of neuroimaging tasks.

\section{\rebuttal{Conclusions}}

\rebuttal{In this work, we performed a thorough assessment of state-of-the-art contrastive learning methods for regression on structural MRI data. Employing a dataset of over 20,000 scans, we show that contrastive learning improves both regression error and transferability compared to standard \rebuttalrev{L1-}supervised deep learning models.}
Scaling contrastive pre-training with the proposed $\mathcal{L}^{exp}$ loss establishes new state-of-the-art performance on brain-age prediction and, crucially, yields embeddings that (i) are robust to scanner variation, (ii) capture accelerated ageing in neurodegenerative disease, and (iii) transfer monotonically to clinical classification tasks.
These properties make CL a strong candidate for building large neuroimaging foundation models that can underpin future precision-medicine pipelines.

\paragraph{Limitations}
First, although we aggregate over 20k scans \rebuttal{and OpenBHB gathers data from different populations}, racial and ethnic diversity remains limited, \rebuttal{especially in the downstream tasks}; future work should verify that the invariance of $\mathcal{L}^{exp}$ extends also to demographic shifts.
Second, our analysis focuses on T1-weighted MRI; multi-modal extensions (e.g., diffusion or functional MRI) could further enrich the learned representations.
Third, we treated the site-prediction task as a proxy for scanner bias; a more comprehensive assessment might include vendor, field strength, and pulse-sequence metadata.

\section*{Acknowledgments}
We acknowledge ISCRA for awarding this project access to the LEONARDO supercomputer, owned by the EuroHPC Joint Undertaking, hosted by CINECA (Italy). This work was granted access to the HPC/AI resources of IDRIS under the allocation 2023-AD011013473R1 made by GENCI.
Data collection and sharing for the Alzheimer's Disease Neuroimaging Initiative (ADNI) is funded by the National Institute on Aging (National Institutes of Health Grant U19AG024904). \rebuttal{This work was implemented on the RAMP~\citep{kegl2018ramp} platform with the support of François Caud at the  DATAIA Institute funded by the ``France 2030'' program ANR-22-CMAS-008.}

\bibliographystyle{model5-names}
\bibliography{refs-curated}

\end{document}